\documentclass[12pt]{iopart}
\usepackage{graphicx}
\begin{document}
\date{}

\title[Letter to the Editor]{A unification in the theory of linearization of
second order nonlinear ordinary differential equations}

\author{V~K~Chandrasekar, M~Senthilvelan and M~Lakshmanan}

\address{Centre for Nonlinear Dynamics, Department of Physics,  Bharathidasan
University, Tiruchirapalli - 620 024, India.}

\begin{abstract} 
In this letter, we introduce a new generalized linearizing transformation (GLT)
for second order nonlinear ordinary differential equations (SNODEs). The well 
known invertible point (IPT) and non-point transformations (NPT) can be derived
as sub-cases of the GLT. A wider class of nonlinear ODEs that cannot be linearized
through NPT and IPT can be linearized by this GLT. We also illustrate how to 
construct GLTs and to identify the form of the linearizable equations and propose a 
procedure to derive the general solution from this GLT for the SNODEs. We
demonstrate the theory with two examples which are of contemporary interest.

\end{abstract}

Linearizing nonlinear ordinary differential equations (NODEs) is still an open 
problem in  the theory of differential equations \cite{olver:1986,Ibragimov1,Steeb}. 
If one raises the
question whether a given arbitrary nonlinear ODE is linearizable or not, no 
definitive
answer can be given in general. Three main points which need attention for further 
understanding of this problem are:
(i) there is still no comprehensive literature available on the types
of transformations that can linearize the ODEs, (ii) the general form of
linearizable equation also differs from transformation to transformation
and (iii) higher order ODEs posses a variety of linearizing
transformations than the lower order ODEs. Due to these reasons no general 
treatment on linearizing
transformations or linearizable equations has been formulated so far. 

In this letter, we make an attempt to unify the linearizing transformations
known for the case of second order nonlinear ODEs (SNODEs) and extend their
scope. As far as the SNODEs
are concerned
it has been shown that, in general, one can 
linearize them through
two different kinds of transformations.  One is the well known invertible point
transformation (IPT) and the other one is the non-point transformation 
(NPT). As far
as the IPT is concerned it has been shown \cite{Steeb,mahomed:1985,
Duarte2,Duarte3,Duarte4,Duarte5} that the most 
general SNODE that can be linearized through 
such a transformation,
\begin{eqnarray} 
X=F(x,t),\quad T=G(t,x), \label {eq01}
\end{eqnarray}
is of the form
\begin{eqnarray} 
\ddot{x}=D(t,x)\dot{x}^3+C(t,x)\dot{x}^2+B(t,x)\dot{x}+A(t,x),
\label {eq02}
\end{eqnarray}
where over dot denotes differentiation with respect to $t$ and 
the functions $A,B,C$ and $D$ should satisfy the following two equations,
\begin{eqnarray} 
\fl \qquad  3D_{tt}+3BD_{t}-3AD_{x}+3DB_{t}+B_{xx}-6DA_{x}+CB_{x}-2CC_{t}
-2C_{tx}=0,\nonumber\\
\fl \qquad  C_{tt}+6AD_{t}-3AC_{x}+3DA_{t}-2B_{tx}-3CA_{x}+3A_{xx}
+2BB_{x}-BC_{t}=0.
\label {eq02a}
\end{eqnarray}
The transformation (\ref{eq01}) converts the equation (\ref{eq02}) into the 
linear `free particle' equation,
\begin{eqnarray} 
\frac{d^2X}{dT^2}=0.\label {eq05a}
\end{eqnarray}

On the other hand, it has also been shown that one can consider NPTs of the form
\begin{eqnarray} 
X=\hat{F}(x,t),\quad dT=\hat{G}(t,x)dt, \label {eq03}
\end{eqnarray}
and linearize the given SNODE. The most general SNODE that can be linearized 
through the transformation (\ref{eq03}) posseses the form \cite{Duarte1}
\begin{eqnarray} 
\ddot{x}+A_2(t,x)\dot{x}^2+A_1(t,x)\dot{x}+A_0(t,x)=0. \label {eq04}
\end{eqnarray}
The set of relations between the functions $A_i$'s, $i=0,1,2,$ and the transformation
(\ref{eq03}) is given by 
\begin{eqnarray} 
A_2=(\hat{G}\hat{F}_{xx}-\hat{F}_x\hat{G}_{x})/K,\nonumber\\
A_1=(2\hat{G}\hat{F}_{xt}-\hat{F}_x\hat{G}_{t}-\hat{F}_t\hat{G}_{x})/K,\nonumber\\
A_0=(\hat{G}\hat{F}_{tt}-\hat{F}_t\hat{G}_{t})/K
\label {eq05}
\end{eqnarray}
with $K=\hat{F}_{x}\hat{G}\neq0$. The NPT also transforms equation (\ref{eq04}) to the free 
particle equation (\ref{eq05a}).
The functions $A_i$'s, $i=0,1,2$, should satisfy the following relations 
\cite{Duarte1}
\begin{eqnarray} 
\fl (i)\qquad \qquad \;\;\; S_1(x,t)=A_{1x}-2A_{2t}=0,\\
S_2(x,t)=2A_{0xx}-2A_{1tx}+2A_0A_{2x}-A_{1x}A_1+2A_{0x}A_2+2A_{2tt}=0.
\end{eqnarray}
\begin{eqnarray}
\fl (ii)\; \mbox{If $S_1(x,t)\neq0$ and $S_2(x,t)\neq0$, then}\nonumber\\ 
S_2^2+2S_{1t}S_2-2S_1^2A_{1t}+4S_1^2A_{0x}+4S_1^2A_0A_2-2S_1S_{2t}
-S_1^2A_1^2=0\\
S_{1x}S_2+S_1^2A_{1x}-2S_1^2A_{2t}-S_1S_{2x}=0.
\end{eqnarray}
The NPT is also called a generalized Sundman transformation, see for example Refs.  
\cite{Euler,Euler2}.

Even though both the IPT and NPT transform the second order nonlinear ODE to 
the free particle equation (\ref{eq05a}), the NPT has some
disadvantages over the former. For example, in the case of IPT one can
unambiguously invert the free particle solution and deduce
the solution of the associated nonlinear
equation, whereas in the case  NPT it is not so straightforward due to the 
non-local
nature of the independent variable.

In this work, we unearth a more general transformation,
\begin{eqnarray} 
X=F(x,t),\quad dT=G(t,x,\dot{x})dt, \label {eq06}
\end{eqnarray} 
and show that this transformation can be utilized to linearize a
wider class of SNODEs and, in particular, certain equations which cannot be 
linearized by the NPT
and IPT. We designate this transformation as the {\it generalized linearizing 
transformation 
(GLT)}. If the function $G$ in (\ref{eq06}) is independent of the variable 
$\dot{x}$ then it becomes an NPT (vide equation (\ref{eq03})). On the other hand $G$ is a perfect 
differentiable function then it becomes an IPT, that is,
$G(t,x,\dot{x})=\frac{d}{dt}\hat{G}(t,x)$, then
$dT=\frac{d\hat{G}}{dt}dt\Rightarrow T=\hat{G}(t,x)$. 
{\it We stress here that (\ref{eq06}) is a unified transformation as it includes
IPT and NPT as special cases}.

We demonstrate our above assertion with the case 
where $G$ is a 
polynomial function in $\dot{x}$ and in particular where it is linear  
in $\dot{x}$ with coefficients which are
arbitrary functions of $t$ and $x$. Indeed, even such a simple case leads to 
interesting results as we see below.
To be specific we focus here on the case
\begin{eqnarray} 
X=F(x,t),\quad dT=(G_1(t,x)\dot{x}+G_2(t,x))dt. \label {eq07}
\end{eqnarray}
Generalizations will be dealt with elsewhere.

Substituting the transformation (\ref{eq07}) into the free particle equation
(\ref{eq05a}), the most general SODE that can be linearized through the GLT
(\ref{eq07}) can be shown to be of the form
\begin{eqnarray} 
\ddot{x}+A_3(t,x)\dot{x}^3+A_2(t,x)\dot{x}^2+A_1(t,x)\dot{x}
+A_0(t,x)=0 \label {eq08}
\end{eqnarray}
and the functions $A_i$'s $i=0,1,2,3$, are connected to the transformation
functions $F$ and $G$
through the relations
\begin{eqnarray} 
A_3=(G_1F_{xx}-F_xG_{1x})/M, \nonumber\\
A_2=(G_2F_{xx}+2G_1F_{xt}-F_xG_{2x}-F_tG_{1x}-F_xG_{1t})/M,\nonumber\\
A_1=(2G_2F_{xt}+G_1F_{tt}-F_xG_{2t}-F_tG_{2x}-F_tG_{1t})/M,\nonumber\\
A_0=(G_2F_{tt}-F_tG_{2t})/M
\label {eq09}
\end{eqnarray}
with $M=F_{x}G_2-F_tG_{1}\neq0$. 

For the given equation one has explicit forms
for the functions $A_i$'s. Now solving equation (\ref{eq09}) with the known
$A_i$'s one can get the linearizing transformation functions $F$ and $G$. Once $F$ and
$G$ are known then using (\ref{eq07}) we can transform (\ref{eq08}) to the 
free particle equation
(\ref{eq05a}) and solving the latter one can get the first integral. 
However, it is difficult to integrate it further unambiguously to obtain the 
general solution due to the non-local nature of the transformation (\ref{eq07}).
We are able to overcome this problem also here and devise a general procedure 
to construct the general solution. In the following we briefly describe the 
procedure. 

Integrating the free particle equation (\ref{eq05a}) once we get
\begin{eqnarray} 
\frac{dX}{dT}=I_1=C(t,x,\dot{x}), \label {eq10}
\end{eqnarray}
where $I_1$ is the first integral. Now rewriting (\ref{eq10}) for $\dot{x}$, 
we get
\begin{eqnarray} 
\dot{x}=f(t,x,I_1), \label {eq11}
\end{eqnarray}
where $f$ is a function of the indicated variables. Due to non-local nature 
of the independent variable we need to consider only a particular solution for 
the free particle equation (\ref{eq05a}), that is, 
\begin{eqnarray} 
X(t,x)=I_1T \label {eq12}
\end{eqnarray}
from which we get
\begin{eqnarray} 
x=g(t,T,I_1), \label {eq13}
\end{eqnarray}
where $g$ is a function of $t,T$ and $I_1$.
Making use of relations (\ref{eq11}) and (\ref{eq13}), equation (\ref{eq07}) 
can be rewritten in the form
\begin{eqnarray} 
dT=h(t,T,I_1)dt, \label {eq13a}
\end{eqnarray}
where again $h$ is a function of $t,T$ and $I_1$. We find that in the case of 
linearizable equations one can separate the variables
$T$ and $t$ in Eq.~(\ref{eq13a}) and integrate the resultant equation which in turn leads to
the general solution.

In the above, we have demonstrated how to deduce linearizing transformation and the
general solution for the given equation. On the other hand one can both 
construct linearizing transformation as well as specific linearizable equations. To
illustrate this let us analyze a particular but important case of 
equation (\ref{eq08}), namely, $A_3=0$ and $A_2=0$ in equation (\ref{eq09}). 
Solving the first
and second equation in (\ref{eq09}) with this restriction, we obtain 
\begin{eqnarray} 
G_1=a(t)F_{x}, \quad 
G_2=a(t)F_{t}-(a_{t}x+b(t))F_{x},
\label {eq18}
\end{eqnarray}
where $a$ and $b$ are arbitrary functions of $t$.
By using equation (\ref{eq18}) in the last two equations in (\ref{eq09}) 
we get 
\begin{eqnarray} 
A_1=S_{x}+\frac{a_t}{(a_{t}x+b)}S+\frac{(a_{tt}x+b_t)}{(a_{t}x+b)},
\label {eq19}\\
A_0=S_{t}+\frac{a_t}{(a_{t}x+b)}S^2+\frac{(a_{tt}x+b_t)}{(a_{t}x+b)}S,
\label {eq20}
\end{eqnarray}
where 
\begin{eqnarray} 
S(x,t)=\frac{F_t}{F_x}.
\label {eq20a}
\end{eqnarray}
Solving equation (\ref{eq19}) we get 
\begin{eqnarray} 
S=\frac{(c(t)-xb_t-\frac{1}{2}x^2a_{tt}+\int A_1(b+xa_t)dx)}{(b+xa_t)},
\label {sol01}
\end{eqnarray}
where $c(t)$ is an arbitrary function of $t$.
Subsituting equation (\ref{sol01}) into (\ref{eq20}) we obtain
\begin{eqnarray} 
\fl \qquad A_0=\frac{a_t(c-xb_t-\frac{1}{2}x^2a_{tt}+\int(b+xa_t)A_1dx)^2}{(b+xa_t)^3}
\nonumber\\
+\frac{c_t-xb_{tt}-\frac{1}{2}x^2a_{ttt}+(\int
((b_t+xa_{tt})A_1+(b+xa_t)A_{1t})dx)}{(b+xa_t)}.
\label {sol02}
\end{eqnarray}
The explicit form of $F$ can be determined by substituting the expression for
$S$ into (\ref{eq20a}) and solving the resultant first order partial differential
equation for $F$. Once $F$ is known $G_1$ and $G_2$ can be fixed using the relation
(\ref{eq18}) which inturn provides us the GLT through (\ref{eq07}). The associated
linearizable equation assumes the form $\ddot{x}+A_1(x,t)\dot{x}+A_0(x,t)=0$,
where $A_0$ is given in equation (\ref{sol02}) and $A_1$ is the given function
in this analysis. 

To illustrate the procedure with a simple but non-trivial example, 
let us consider the case $A_1=kx^q$, where $k$ and $q$ are arbitrary
parameters, and fix the arbitrary functions
$a,b$ and $c$ such as $a(t)=t,\;b(t)=c(t)=0$, so that the equation (\ref{sol01}) gives us
\begin{eqnarray} 
S=\frac{k}{(q+2)}x^{q+1}.\label {eq21b}
\end{eqnarray}
Once $S$ is known $F$ and $A_0$ can be fixed through the relations 
(\ref{eq20a}) and (\ref{sol02}) of the form
\begin{eqnarray} 
A_0=\frac{k^2}{(q+2)^2}x^{2q+1}\quad \mbox{and}\quad 
F=\frac{k}{q+2}t-\frac{1}{qx^q}.
\label {eq21a}
\end{eqnarray}
The forms of $A_0$ and $A_1$ fix the linearizable equation (\ref{eq08}) to the 
form 
\begin{eqnarray} 
\ddot{x}+kx^q\dot{x}+\frac{k^2}{(q+2)^2}x^{2q+1}=0. \label {eq14}
\end{eqnarray}
Since $a(t)=t$ and $b(t)=0$, from (\ref{eq18}) we have 
\begin{eqnarray} 
G_1=\frac{t}{x^{q+1}},\qquad  G_2=\frac{kt}{q+2}-\frac{1}{x^q}.\label {eq14a}
\end{eqnarray} 
As a consequence the linearizing
transformation turns out to be
\begin{eqnarray} 
X=\frac{k}{q+2}t-\frac{1}{qx^q},\qquad 
dT=\bigg[-t\bigg(\frac{k}{q+2}+\frac{\dot{x}}{x^{q+1}}\bigg)
+\frac{1}{x^q}\bigg]dt.
 \label {eq22}
\end{eqnarray}
It is easy to check that equation (\ref{eq14}) can be linearized to the free
particle equation (\ref{eq05a}) through the transformation (\ref{eq22}).

Equation (\ref{eq14}) and its sub-cases have been widely discussed in the contemporary
literature. In particular, Mahomed and Leach \cite{mahomed:1985} have shown that
equation (\ref{eq14}) with $q=1$ is one of the SNODEs that can be linearized to
the free particle equation (\ref{eq05a}) through the IPT 
$X=\frac{k}{3}t-\frac{1}{x}$ and $T=\frac{t}{x}-\frac{kt^2}{6}$. Consequently, the group
invariance and integrability properties of this sub-case, namely, $q=1$, and the
general equation (\ref{eq14}) have been studied extensively by different authors,
see for example Refs. \cite{Davis,feix:1997,leach:1988a,Chand1}. However, in the literature 
equation
(\ref{eq14}) has been shown to be linearizable to free particle equation only
for the value $q=1$. For other values of $q$ the linearization of this equation
through IPT or NPT was not known. {\it But in the present work we have proved
above that one can
linearize the entire class of equation (\ref{eq14})
under the one general transformation (\ref{eq22}), irrespective of the
value of $q$}. One may note that
choosing $q=1$ the GLT (\ref{eq22}) coincides exactly with the point
transformation for equation (\ref{eq14}) with the same parametric
restriction. This example further confirms the arguments
that IPT is a sub-case of GLT. 

In the following 
we derive the general solution of (\ref{eq14}) using our procedure discussed 
through the equations (\ref{eq10})-(\ref{eq13a}).
Using (\ref{eq22}) in equation (\ref{eq10}), we obtain the first integral in
the form
\begin{eqnarray} 
I_1=\frac{(\frac{k}{q+2}x^{q+1}+\dot{x})}
{-t(\frac{k}{q+2}x^{q+1}+\dot{x})+x}. \label {eq23}
\end{eqnarray}
Rewriting (\ref{eq23}) for $\dot{x}$, we get
\begin{eqnarray} 
 \dot{x}=-\frac{k}{q+2}x^{q+1}+\frac{I_1}{(1+I_1t)}x.
\label {eq25}
\end{eqnarray}
Making use of the particular solution for the free particle equation given
in equation (\ref{eq12}) and rewriting this for $x$ in equation (\ref{eq22}), 
we get
\begin{eqnarray} 
x=(\frac{1}{q(\frac{k}{q+2}t-I_1T)})^{\frac{1}{q}}. 
\label {eq25a}
\end{eqnarray}
Substituting (\ref{eq25}) and (\ref{eq25a}) in the second equation in
(\ref{eq22}), we obtain
\begin{eqnarray} 
dT=q\bigg(\frac{1}{1+I_1t}\bigg)\bigg(\frac{k}{q+2}t-I_1T\bigg)dt. \label {eq26}
\end{eqnarray}
Rewriting equation (\ref{eq26}) in the form
\begin{eqnarray} 
\frac{dT}{dt}+\frac{qI_1}{1+I_1t}T=\frac{kq}{q+2}(\frac{t}{1+I_1t}) 
\label {eq26a}
\end{eqnarray} 
and integrating the resultant equation, (\ref{eq26a}), we get
\begin{eqnarray} 
T=(1+I_1t)^{-q}\bigg(I_2+\frac{(1+I_1t)^{q}(qI_1t-1)}{I_1^2(2+3q+q^2)}\bigg), 
\label {eq26b}
\end{eqnarray}
where $I_2$ is the second integration constant. Substituting the resultant 
expression for $T$ into (\ref{eq25a}) we obtain the general solution of 
(\ref{eq14}), that is,
\begin{eqnarray} 
x(t)=\bigg(\frac{I_1(q+1)(q+2)(1+I_1t)^q}
{q(k(1+I_1t)^{q+1}-I_1^2I_2(2+3q+q^2))}\bigg)^{\frac{1}{q}},\label {eq27}
\end{eqnarray}
which is the same as the one obtained by Feix et al. \cite{feix:1997}.

Equation (\ref{eq14}) is not an isolated example that can be linearized through
the GLT.
In fact, one can linearize a larger class of equations through this GLT and obtain the
general solution. This is mainly due to the presence of arbitrary functions,
namely, $a(t),\;b(t)$ and $c(t)$ in the determining
equations for given $A_1(x,t)$. To demounstrate this, we consider a slightly  
more general form 
\begin{eqnarray}
A_1=k_1x^q+k_2,\quad A_3=A_2=0,\label {eq28}
\end{eqnarray}
where $k_1,k_2$ and $q$ are arbitrary constants, in equation (\ref{eq08}). 
In the present example we have
included an additive constant, $k_2$, in the function $A_1$ and left the
other two functions $A_2$ and $A_3$ the same as before. However, this 
additive constant itself enlarges the class of linearizable equations 
considerably, as we see below. 

Let us again fix the arbitrary functions $a,b$ and
$c$ of the same form as in the previous example, that is, $a(t)=t,\;b(t)=0$ and
$c(t)=0$, so that we get $S=\frac{k_2}{2}x+\frac{k_1}{q+2}x^{q+1}$. The
respective linearizable equation turns out to 
be
\begin{eqnarray}
\ddot{x}+(k_1x^q+k_2)\dot{x}+\frac{k_1^2}{(q+2)^2}x^{2q+1}
+\frac{k_1k_2}{q+2}x^{q+1}+\frac{k_2^2}{4} x=0 \label {eq30a}
\end{eqnarray}
and the GLT becomes
\begin{eqnarray} 
&X&=\frac{2k_1}{q(q+1)k_2}-\bigg(\frac{1}{qx^q}+\frac{2k_1}{q(q+1)k_2}\bigg)
e^{-\frac{q}{2}k_2t},\nonumber\\
&dT&=\bigg[\frac{1}{x^q}(1-\frac{k_2}{2}t)
-t(\frac{\dot{x}}{x^{q+1}}+\frac{k_1}{(q+1)})\bigg]e^{-\frac{q}{2}k_2t}dt.
 \label {eq32a} 
\end{eqnarray}
One may note that in the limit $k_2\rightarrow0$ both the 
linearizing transformations, (\ref{eq32a}) and
the linearizable equation, (\ref{eq30a}), reduce to the earlier example (vide 
equations (\ref{eq22}) and (\ref{eq14}) respectively).

The associated first integral reads
\begin{eqnarray} 
I_1=\frac{dX}{dT}=\frac{(\frac{k_2}{2}x+\frac{k_1}{q+2}x^{q+1}+\dot{x})}
{-t(\frac{k_2}{2}x+\frac{k_1}{q+2}x^{q+1}+\dot{x})+x}.
\label {eq33a}
\end{eqnarray}
Repeating the same steps given in the previous example one can get the general 
solution for
the equation (\ref{eq30a}) in the form
\begin{eqnarray} 
 x(t)=(I_1+t)e^{-\frac{k_2}{2}t}\bigg(I_2+\frac{qk_1}{(q+2)}
 \int_0^t{e^{-\frac{qk_2}{2}t'}(I_1+t')^q dt'}\bigg)^{-\frac{1}{q}},
 \label{eq34a}
\end{eqnarray}
where $I_2$ is the second integration constant.

Next we choose the arbitrary function $a(t)$ in an exponential form,
namely, $a(t)=e^{\alpha t}$, where $\alpha$ is a constant, with $b(t)=c(t)=0$. 
In this case we get 
\begin{eqnarray}
S=\bigg(\frac{k_2+\alpha}{2}\bigg)x+\frac{k_1}{q+2}x^{q+1} 
\quad \mbox{and}\quad \alpha=\sqrt{k_2^2-4\lambda},
\label {eq29a}
\end{eqnarray}
where $\lambda$ is an arbitrary parameter. The functions $b,c$ and $A_1$ give
$A_0$ through the relation (\ref{sol02}) which in turn gives us a new linearizable 
equation of the form 
\begin{eqnarray}
\ddot{x}+(k_1x^q+k_2)\dot{x}+\frac{k_1^2}{(q+2)^2}x^{2q+1}
+\frac{k_1k_2}{(q+2)}x^{q+1}+\lambda x=0. \label {eq30}
\end{eqnarray}
Proceeding further we obtain the GLT in the form 
\begin{eqnarray} 
&X&=\bigg(\frac{(\alpha-k_2)}{2q\lambda}-\frac{(q+2)}{qk_1x^q}\bigg)
e^{-\frac{q}{2}(k_2+\alpha)t},\nonumber\\
&dT&=\bigg[\frac{(q+2)}{k_1x^{q+1}}\dot{x}+\frac{(q+2)}{2k_1x^q}(k_2-\alpha)
+1\bigg]e^{-(\frac{q}{2}k_2+\frac{(q-2)}{2}\alpha)t}dt.
 \label {eq32}
\end{eqnarray}
Now one can check that the transformations (\ref{eq32a}) and (\ref{eq32}) 
transform (\ref{eq30a}) and (\ref{eq30}) to
the free particle equation (\ref{eq05a}). The first integrals for  
equation (\ref{eq32}) can be constructed of the form
\begin{eqnarray} 
I_1=\frac{dX}{dT}=e^{-\alpha t}
\left(\frac{\dot{x}+\frac{(k_2+\alpha)}{2}x+\frac{k_1}{q+2}x^{q+1}}
{\dot{x}+\frac{(k_2-\alpha)}{2}x+\frac{k_1}{q+2}x^{q+1}}\right),
\label {eq33}
\end{eqnarray}
and the general solution take the form
\begin{eqnarray} 
x(t)=\bigg(e^{\alpha t}-I_1\bigg)e^{-\frac{1}{2}(k_2+\alpha)t}
\bigg(I_2+\frac{qk_1}{(q+2)}\int_0^t{\bigg(\frac{e^{\alpha t'}-I_1}
{e^{\frac{1}{2}(k_2+\alpha)t'}}\bigg)^q dt'}\bigg)
^{-\frac{1}{q}},\label{eq34}
\end{eqnarray}
where $I_2$ is the second integration constant. Equations (\ref{eq34a}) and
(\ref{eq34}) can be integrated further explicitly using the standard method
\cite{Gradshteyn}. To our knowledge the solutions (\ref{eq34a}) and (\ref{eq34}) are new
to the literature.

We note that in the case $q=1$, the terms on the right hand side in the second
equation in (\ref{eq32}) can be written as a perfect 
derivative term and consequently leads us to the same IPT for the equation
(\ref{eq30}) respectively with $q=1$ obtained in 
Refs. \cite{Chand1,Chand2}.

In this paper, we have introduced a new generalized linearizing transformation
which can be used to linearize a class of equations that cannot be linearized
by either IPT or NPT. In fact both IPT and NPT can be derived as sub-cases from
the proposed GLT. Since the independent variable is in a non-local form in the
GLT, we have devised an algorithm to rewrite the new variables in terms of old 
variables. Needless
to say, this algorithm can also be used in the case of NPT also. Importantly,
we have illustrated our theory with certain concrete examples which are of
contemporary interest. Naturally one can also construct GLTs involving more
general forms of $\dot{x}$ in (\ref{eq06}) and identify new linearizable equations. 
The procedure can also be extended to higher order ODEs. The details will
be discussed separately.

The work of VKC is supported by CSIR in the form of a CSIR Senior Research
Fellowship.  The work of MS and ML forms part of a Department of Science and 
Technology, Government of India sponsored research project.

\end{document}